\documentclass[12pt,preprint]{aastex}

\input epsf.sty

\newcommand{\etal}{et~al.\ }
\newcommand{\eg}{e.g.\ }

\newcommand{\Msun}{M_{\odot}}

\newcommand{\kms}{km~s$^{-1}$}

\newcommand{\CII}{C~{\sc ii}}

\newcommand{\SiII}{Si~{\sc ii}}

\newcommand{\CaII}{Ca~{\sc ii}}
\newcommand{\TiII}{Ti~{\sc ii}}

\newcommand{\FeII}{Fe~{\sc ii}}
\newcommand{\FeIII}{Fe~{\sc iii}}

\begin{document}

\title{High-Velocity Features: a ubiquitous property of Type Ia SNe}

\author{P.A.~Mazzali\altaffilmark{1,2}, S.~Benetti\altaffilmark{3},
G.~Altavilla\altaffilmark{4}, G.~Blanc\altaffilmark{3}, 
E.~Cappellaro\altaffilmark{5}, N.~Elias-Rosa\altaffilmark{3}, 
G.~Garavini\altaffilmark{6,7}, A.~Goobar\altaffilmark{7},
A.~Harutyunyan\altaffilmark{3}, R.~Kotak\altaffilmark{8},
B.~Leibundgut\altaffilmark{9}, P.~Lundqvist\altaffilmark{10},
S.~Mattila\altaffilmark{10}, J.~Mendez\altaffilmark{4}, 
S.~Nobili\altaffilmark{6,7}, R.~Pain\altaffilmark{6},
A.~Pastorello\altaffilmark{3,2}, 
F.~Patat\altaffilmark{9}, G.~Pignata\altaffilmark{9}, 
Ph.~Podsiadlowski\altaffilmark{11}, P.~Ruiz-Lapuente\altaffilmark{4}, 
M.~Salvo\altaffilmark{12}, B.P.~Schmidt\altaffilmark{12},  
J.~Sollerman\altaffilmark{10}, V.~Stanishev\altaffilmark{7}, 
M.~Stehle\altaffilmark{2}, C.~Tout\altaffilmark{13},
M.~Turatto\altaffilmark{3}, and W.~Hillebrandt\altaffilmark{2}
}

\altaffiltext{1}{INAF-Osserv.\ Astron., Via Tiepolo, 11,
  34131 Trieste, Italy}
\altaffiltext{2}{Max-Planck Institut f\"ur Astrophysik, 
  Karl-Schwarzschildstr. 1, 85748 Garching, Germany}
\altaffiltext{3}{INAF-Osserv.\ Astron.\ di Padova, 35122 Padova, Italy}
\altaffiltext{4}{Department of Astronomy, U. Barcelona, Barcelona, Spain}
\altaffiltext{5}{INAF-Osserv.\ Astron.\ di Capodimonte, Napoli, Italy}
\altaffiltext{6}{Lab. de Physique Nucl\'{e}aire et de Haute Energies,
  CNRS-IN2P3, Univ. Paris VI and VII, Paris, France}
\altaffiltext{7}{Dept. of Physics, Stockholm Univ.,  
  SE-106 91 Stockholm, Sweden}
\altaffiltext{8}{Blackett Laboratory-Imperial College, London, England}
\altaffiltext{9}{European Southern Observatory, 85748 Garching, Germany}
\altaffiltext{10}{Stockholm Obs., SE-106 91 Stockholm, Sweden}
\altaffiltext{11}{Dept. of Astronomy, Oxford University, Oxford, England}
\altaffiltext{12}{Mt.Stromlo and Siding Spring Obs., Canberra, Australia}
\altaffiltext{13}{Institute of Astronomy, Cambridge, England}

\begin{abstract}

Evidence of high-velocity features such as those seen in the near-maximum
spectra of some Type Ia Supernovae (\eg SN~2000cx) has been searched for in the
available SN~Ia spectra observed earlier than one week before $B$ maximum. 
Recent observational efforts have doubled the number of SNe~Ia with very early
spectra.  Remarkably, all SNe~Ia with early data (7 in our RTN sample and 10
from other programmes) show signs of such features, to a greater or lesser
degree, in \CaII\ IR, and some also in \SiII\ $\lambda\lambda$\,6255\AA\ line. 
High-velocity features may be interpreted as abundance or density enhancements. 
Abundance enhancements would imply an outer region dominated by Si and Ca. 
Density enhancements may result from the sweeping up of circumstellar material
by the highest velocity SN ejecta. In this scenario, the high incidence of HVFs
suggests that a thick disc and/or a high-density companion wind surrounds the
exploding white dwarf, as may be the case in Single Degenerate systems. 
Large-scale angular fluctuations in the radial density and abundance
distribution may also be responsible: this could originate in the explosion, and
would suggest a deflagration as the more likely explosion mechanism.  
CSM-interaction and surface fluctuations may coexist, possibly leaving different
signatures on the spectrum.  In some SNe the HVFs are narrowly confined in
velocity, suggesting the ejection of blobs of burned material.

\end{abstract}


\section{Introduction}

The importance of Type Ia Supernovae (SNe~Ia) for cosmological studies has
motivated increased efforts to understand how these bright explosions work and
how their observational properties depend on the physical processes involved.
Three-dimensional modelling of the thermonuclear explosion of a CO White Dwarf
reaching the Chandrasekhar limit holds the promise to describe successfully how
SNe~Ia are born.  On the observational side, several groups are trying to follow
nearby SNe~Ia with an increased temporal coverage, so that explosion models can
be tested and the range of observational SN properties fully described.

In particular, it is now agreed that it is important to observe SNe~Ia as early
as possible.  Early observations make it easier to distinguish SNe~Ia from
SNe~Ic, and allow firmer constraints to be placed on the actual  explosion date
\citep[\eg][]{rie99, MS00}, which is important for the use of SNe~Ia in
cosmology.  Moreover, early spectra explore the outer layers of the SN ejecta. 
These layers may hold traces of the properties of the progenitor system, whose
nature may thereby be revealed \citep[\eg][]{len00}.

As the number of high S/N observations increases, a few cases have been reported
of SNe~Ia with overall normal properties, but displaying high-velocity
absorption lines in particular of the \CaII\ IR triplet.  The first such 
features identified were \CaII\ and \FeII\ lines in SN~1994D \citep{hat99}. 
Further evidence came from SN~2000cx \citep{li01}, which showed two strong sets
of \CaII\ IR lines, at different high velocities, in addition to the
photospheric component.  High-velocity features were also observed in SN~2001el
\citep{wang03}, SN~2003du \citep{ger04}, and in SN~1999ee.  For SN~1999ee,
\citet{maz05} note that high velocity absorption is present not only in \CaII\
IR, but also in the strong \SiII\ 6355\AA\ line. High-velocity features (HVFs)
may be indicative of interaction with a circumstellar medium \citep{ger04}, of
an intrinsic three-dimensional structure of the explosion \citep[\eg][]{maz05},
or possibly of a combination of these two situations. 

Available evidence suggests that HVFs are more common before maximum. To obtain
early coverage of nearby SNe~Ia is one of the aims of an EU-funded RTN, linking
most European researchers working on SNe~Ia.  The RTN has collected data for a
number of SNe~Ia, 7 of which have spectra earlier than one week before $B$
maximum.

In this paper, we review all spectra earlier than $-7$ days to look for
evidence  of HVFs and assess the ubiquity of this phenomenon.  We define a HVF
any absorption that is at a higher velocity than the corresponding photospheric
line.  A detached HVF is one that is not blended with the photospheric line, and
is likely to be due to material that is physically detached from the
photospheric region. In Sect.\ 2 we present the 7 RTN SNe with early spectra,
and in Sect.\ 3 we discuss the existing non-RTN sample. Sect. 4 contains a
discussion of the frequency of HVFs and the physical hypotheses we can make
about their ubiquity.

\section{High-velocity features in the the RTN objects}

Figure 1 shows spectra of the 7 RTN SNe~Ia with early observations. \CaII\ IR
HVFs are evident in many SNe, sometimes detached from the photospheric
component, and are typically stronger in earlier spectra (\eg SNe~2002dj and
2003du).  Figure 2 shows the \CaII\ IR profiles.  Table 1 lists the results of
fitting the profiles with two separate gaussians, representing the two 
components, following \citet{maz05}. The last column of Table~1 is the ratio of
the intensities of the two gaussians. HVFs are not common in the \SiII\ line. 
In some objects, however, the shape and position of the line suddenly change,
indicating the presence and progressive thinning out of a HVF. Below we discuss
each SN in turn.

{\bf SN~2001el.}  \citet{wang03} observed strong, detached \CaII\ IR at $v \sim
22-26000$\,\kms\ between $-4$ and $+2$ days, accompanied by significant line
polarisation.  They interpreted this as an indication of the presence of a clump
of material at $v \sim 20-26000$\,\kms.  The HVF was also observed in an
earlier spectrum \citep[day $-9$,][]{mat05}.  The two strongest lines in the
triplet are not blended in the HVF.  A similar situation occurred in SN~1999ee
(Sect.\ 3). The two lines are also partially unblended in the photospheric
component in the near-maximum spectrum, as in SNe 2003du and 2003cg. The
flat-bottomed \SiII\ profile at day $-9$ \citep{mat05} could be a HVF, as in
SN~1990N (Sect.\ 3).

{\bf SN~2002bo.}  \citet{ben04} published the spectra of this normal SN~Ia.  The
earliest spectrum covering \CaII\ IR is at $-8.5$ days. Although a detached
feature is not visible, the line appears blue and broad, becoming narrower and
shifting to the red at later epochs. The profile can be decomposed into a
high-velocity component near 22000\,\kms\ and a photospheric one with similar
strength near 15000\,\kms.  \citet{ste05} reproduce the profile evolution using
a density enhancement at $v \sim 20000$\,\kms.  As the ejecta expand the density
enhancement becomes optically thin, and by day $-3.6$ the HVF is very weak. The
\SiII\ line profile also changes noticeably, losing strength in the blue between
day $-11$ and maximum, possibly indicative of a HVF.

{\bf SN~2002dj.}  The earliest red spectrum was observed at $-11$ days.  The
\CaII\ IR profile is extremely blue, with central velocity $v \sim
27600$\,\kms.  The HVF is however not detached: the photospheric component,
which has a similar strength, is at a high velocity, as might be expected since
the epoch of the spectrum is rather early.  The photospheric component becomes
dominant as the HVF fades: by day $-3$ the HVF seems to have disappeared.

{\bf SN~2002er.}  The earliest \CaII\ IR spectrum is only at $-7.4$ days.  The
line is weaker than in other SNe, but it is very broad.  The weakness of both
\CaII\ IR and H\&K may indicate a low Ca abundance or a high ionisation degree.
Nevertheless, the \CaII\ IR triplet can be decomposed into a HVF ($v \sim
23000$\,\kms) and a stronger photospheric component ($v \sim 15000$\,\kms).  
The HVF has practically disappeared in the spectrum near maximum.

{\bf SN~2003cg.}  The earliest spectrum covering \CaII\ IR is at $-8.5$ days.
All \CaII\ lines are extremely weak in this SN, deserving some investigation.  A
weak HVF is visible in the \CaII\ IR triplet at $v \sim 22000$\,\kms.  The HVF
is detached owing not so much to its velocity but rather to the weakness of the
photospheric component (near 12000\,\kms), where the two strongest lines of the
triplet are unblended.  The HVF disappears by the time of maximum.

{\bf SN~2003du.}  \citet{ger04} measured a HVF at $v \sim 18000$\,\kms, and
suggest it is due to the accumulation of CSM.  Our first \CaII\ IR spectrum is
very early, $-11.5$ days, and it shows a very strong HVF, detached at $v \sim
22500$\,\kms.  The two strongest lines are not completely blended, indicating
some degree of velocity confinement. The photospheric component is weak at
first, but dominates on day $+0.6$, when a weak HVF may still be present: at
this epoch the profile resembles SN~2003cg. The \SiII\ line evolves as in
SN~2002bo.

{\bf SN~2003kf.} In the earliest spectrum ($-9$ days) \CaII\ IR shows a strong, 
detached HVF at $v \sim 23500$\,\kms, and a weak photospheric component. On day
$-5$, however, the two components have a similar strength. 

In summary, all 7 RTN SNe with early spectra show HVFs in \CaII\ IR; SNe~2001el,
2002bo and 2003du also in the \SiII\ line.  \CaII\ IR HVFs are detached if the
photospheric component is weak, otherwise they broaden the photospheric
absorption to the blue.  \SiII\ HVFs are much weaker, and never detached. HVFs
lose strength with time: in most cases their presence near maximum is doubtful.

\section{Early observations of other SNe~Ia} 

Several other SNe~Ia have been observed at very early times by different
observers in the last 20 years. Here we review the available data and discuss
the evidence for HVFs. 

{\bf SN~1984A.} The very high \SiII\ velocities of this SN, reaching  $\sim
16500$\,\kms\ \citep{ben05}, may be indicative of a HVF. Spectral coverage 
starts as early as $-7$ days \citep{bar89}, but unfortunately the \CaII\ IR
region is not covered.  

{\bf SN~1990N.}  The earliest spectrum \citep[$-14$ days][]{lei91} does not
cover \CaII\ IR fully. An absorption near 8000\,\AA\ may be a HVF.  The flat
absorption trough of the \SiII\ line, however, has been the subject of debate. 
\citet{fish97} suggested that it was due to high-velocity \CII.  \citet{maz00}
showed that high-velocity \SiII\ is a more physical possibility. A similar
\SiII\ profile was observed in SNe~2001el and 1999ee (see below).  

{\bf SN~1991T.}  This peculiar object, the prototype of a subclass of SNe~Ia,
has very early spectra (starting at $-12$ days) that extend well into the red
\citep{rlp92}.  However, the ionisation conditions must have been so high in the
pre-maximum phase that neither \CaII\ nor \SiII\ lines were visible
\citep{maz95}.  The only strong lines at very early epochs were \FeIII, and
these did not show clear signs of HVFs.

{\bf SN~1992A.}  The red part of the earliest spectrum published by
\citet{kir93} was taken 7 days before maximum.  The blue absorption wing of
\CaII\ IR is not reproduced by the synthetic spectra and may be due to a HVF at
$\sim 22000$\,\kms.  According to \citet{kir93}, Si extends out to at least
19000\,\kms.

{\bf SN~1994D.}  HVFs were noticed by \citet{hat99} at day $-8.9$. \citet{hat99}
showed through spectral modelling that \CaII\ and \FeII\ are present at $25000 <
v < 40000$\,\kms, detached from the photospheric component ($v < 16000$\,\kms). 
The \CaII\ IR profile is very similar to that of SN~2003kf at $-9$ days.  The
rapid fading of the HVF in SN~2003kf suggests that in both SNe HVFs may have
been much stronger earlier on.  

{\bf SN~1998bu.}  The earliest spectrum \citep[$-6.8$ days][]{her00} shows a
weak \CaII\ IR detached feature at $v \sim 20000$\,\kms, which may be weakly
present until maximum.  

{\bf SN~1999aa.} This very slow-declining SN was spectroscopically similar to
SN~1991T at very early times, and then rapidly transitioned to normal-looking
spectra. The earliest spectrum ($-11$ days) showed only very few lines, mostly
of \FeIII. However, \citet{gar04}, based on spectral modelling, suggest the
presence of HVFs in both \CaII\ and \CII\ lines at $v > 20000$\,\kms\ in the
spectra at $-$7, $-3$ and $-1$ days. The time evolution of the HVFs may be
consistent with \CaII\ forming a clump in the outermost layers of the ejecta.

{\bf SN~1999ac.}  The pre-maximum spectra of this SN do not show HVFs.  In
particular, \CaII\ IR at $-15$ days has a minimum at 15500\,\kms\ and a blue
edge at 30000\,\kms, which may correspond to the photospheric component. 
However, SN~1999ac is a peculiar object in many respects.  Its
spectro-photometric characteristics do not match those of any of the known
supernovae (Garavini et al 2005, in preparation).

{\bf SN~1999ee.}  \citet{maz05} noticed HVFs in both \CaII\ IR and \SiII\
6355\AA, at $v \sim 18-20000$\,\kms, from $-9$ to $+3$ days.  The two strongest
\CaII\ IR lines are not blended in the HVF, indicating a narrow velocity
confinement of the material, and remain at an almost constant velocity until
they disappear.  \citet{maz05} modelled the HVFs using a high-velocity density
enhancement, since an abundance enhancement requires that the outer layers are
composed only of Si and Ca.  They find that adding $\sim 0.1 \Msun$ of
SN-composition material can explain the observed HVFs, but argue that an outer
H-rich layer containing $\sim 0.004 \Msun$ gives an improved solution.  HVFs due
to SN-composition material would be attributed to a property of the explosion,
but the accumulation of H-rich material would necessarily be due to CSM
interaction.  Both situations may exist in SN~1999ee, the H-rich CSM giving rise
to line broadening, and SN ejecta to the narrow HVFs.

{\bf SN~2000cx.}  This very peculiar SN shows two very strong sets of detached
\CaII\ IR at different velocities \citep[][Lundqvist \etal 2005, in prep.]
{li01}.  There is no evidence of HVFs in \SiII\ 6355\AA, but \citet{bra04}
report the presence of \TiII\ HVFs.  \citet{tho04} infer the presence of a
detached \CaII\ shell at $v > 16000$\,\kms.   

\section{Discussion} 

HVFs are present, at least in \CaII\ IR, in almost all SNe~Ia with early
spectra, but they have different strength and duration. Some SNe also show
\SiII\ HVFs. Spectroscopically, HVFs could be due to abundance or density
enhancements.

An abundance enhancement would be a property of the explosion. However,
\citet{maz05} show that even\ extreme abundance enhancements, such that Si and 
Ca dominate the outer regions of the ejecta, cannot produce sufficiently strong
HVFs. That was a 1D test for SN~1999ee, where the HVFs are not the strongest. 3D
models do show fingers of burned material penetrating to high velocities, but
then this material has a covering factor $< 1$. Therefore, abundance
enhancements alone can only partially explain the observations at best, and
density enhancements must also be invoked.  Density enhancements would increase
the CaII\ and \SiII\ line opacity not only directly, but also indirectly through
recombination in the HVFs (Ca and Si are typically triply ionised in the
outemost parts of the ejecta). 

Density enhancements may themselves be a property of the explosion: large-scale
angular fluctuations, similar to the fingers of burned material in a
deflagration, density blobs offering different covering factors depending on the
orientation, or the effect of the onset of a delayed detonation.
The fact that polarisation was observed in coincidence with the HVFs in 
SN~2001el \citep{wang03} indicates that HVFs are due to 3D effects.  

Alternatively, density enhancements may be the result of the interaction of the
SN ejecta with CSM. In this case, constraints can be placed on the nature of 
the CSM itself. First, the CSM must offer a large angular size to the ejecta,
otherwise orientation effects would imply that only some SNe should show HVFs. 
Secondly, any CSM must be located near the SN for it to affect the spectrum
immediately after the explosion.  
\footnote{ The strongly interacting SN~Ia 2002ic \citep{ham03} is an outlier, 
since interaction starts only after maximum.  A different progenitor system may
be involved in this case (\eg an AGB companion). }

The CSM closest to the WD is the accretion disc. If the disc is responsible for
the HVFs, it must be thick compared to the size of the WD \citep{ger04}.
Variability in strength and duration of the interaction may depend on both disc
size and viewing angle.
Alternatively, the CSM may be a wind emanating from either star in the system,
the strength of the interaction depending on the wind properties.  However, for
SN~1999ee \citet{maz05} find that as much as $0.004 \Msun$ of H-rich material
may have accumulated on top of the SN ejecta, and an even larger value was
obtained by \citet{ger04} for SN~2003du.  The implied wind mass loss rate seems
unreasonably high for a fast WD wind ($\sim 10^{-2} \Msun$\,yr$^{-1}$), but it
is high also for a slow red giant wind ($\sim 10^{-4} \Msun$\,yr$^{-1}$).  Since
the material must surround the system, one possibilty is enhanced mass loss from
the system through the $L_1$ point.  This should take place at a much higher
rate than in single stars.

One property that would distinguish a density enhancement resulting from the
explosion from one due to CSM is the composition of the added material.
\citet{maz05} show that the narrow, unblended \CaII\ IR HVFs in SN~1999ee can be
reproduced with a density enhancement of SN-composition material, but that
H-rich material helps to reproduce the overall line width, so this may be a case
where the two effects coexist. Unblended HVFs are also observed in SN~2001el.
The case of SN~2000cx, with a double set of HVFs, is also interesting. 
Possibly, the unblended HVFs are due to blobs of material generated in the
explosion, while broad HVFs are the result of CSM accretion.  If the 3D
structure of the ejecta is such that the density profile varies with direction,
it may be expected that the range of velocities of the swept-up CSM could be
broader than that of the blobs. A study of the composition of the CSM could
reveal the nature of SNe~Ia progenitors.

Early and frequent observations, both spectroscopic and polarimetric, and a
study of the correlation between HVF strength and polarization \citep[\eg
SNe~2001el, 2004dt,][resp.]{wang03,wang04}, may reveal what the HVFs imply.


This work was partly supported by the European Research and Training Network
2002-2006 "The Physics of Type Ia Supernovae" (contract  HPRN-CT-2002-00303).




\clearpage


\begin{deluxetable}{ccccccc}
\tabletypesize{\scriptsize}
\tablecaption{Properties of the \CaII\ IR triplet }
\tablewidth{0pt}
\tablehead{ 
\colhead{SN} & 
\colhead{day} &
\colhead{$v(ph)$} &
\colhead{$W(ph)$} &
\colhead{$v(hv)$} &
\colhead{$W(hv)$} &
\colhead{$\frac{F(ph)}{F(hv)}$} \\
}
\startdata 
2002dj & -11  &  17700 &  11200 &  27600 &   9100  &  1.2 \\
       &  -3  &  14900 &  11200 &  21800 &   7000  &   17 \\
       &      &        &        &        &         &	  \\
2001el &  -9  &  17100 &  14000 &  23800 &   7700  &  0.9 \\
       &  +1  &  11800 &   8700 &  21400 &   7000  &  0.8 \\
       &      &        &        &        &         &	  \\
2003du & -11  &  15500 &  14000 &  22500 &   6600  &  1.3 \\
       & +0.6 &  10400 &   8700 &  19300 &   4500  &    4 \\
       &      &        &        &        &         &	  \\
2003kf &  -9  &  12600 &  14000 &  23500 &   8700  &  0.6 \\
       &  -5  &  11600 &   9800 &  22100 &   8000  &  0.9 \\
       &      &        &        &        &         &	  \\
2002er &  -7  &  15600 &  12200 &  23100 &   6600  &  2.3 \\
       &   0  &  13100 &  10500 &  20200 &   6600  &   17 \\
       &      &        &        &        &         &	  \\
2002bo &  -8  &  14900 &  10500 &  22100 &   8000  &  1.6 \\
       &  0.1 &  13100 &   9800 &  18600 &   5600  &    5 \\
       &      &        &        &        &         &	  \\
2003cg & -8.5 &  12700 &  10100 &  22000 &   4900  &  4.2 \\
       & -0.6 &  11700 &   8700 &  19400 &   3500  &   11 \\
\enddata 
\tablecomments{Velocities (\kms) are relative to the average \CaII\ IR 
triplet wavelength, 8579.1\AA, and rounded to the nearest 100. Measurements 
are more uncertain when the two components are blended (\eg SN~2002dj).  } 
\end{deluxetable}


\newpage
\begin{figure*}
\plotone{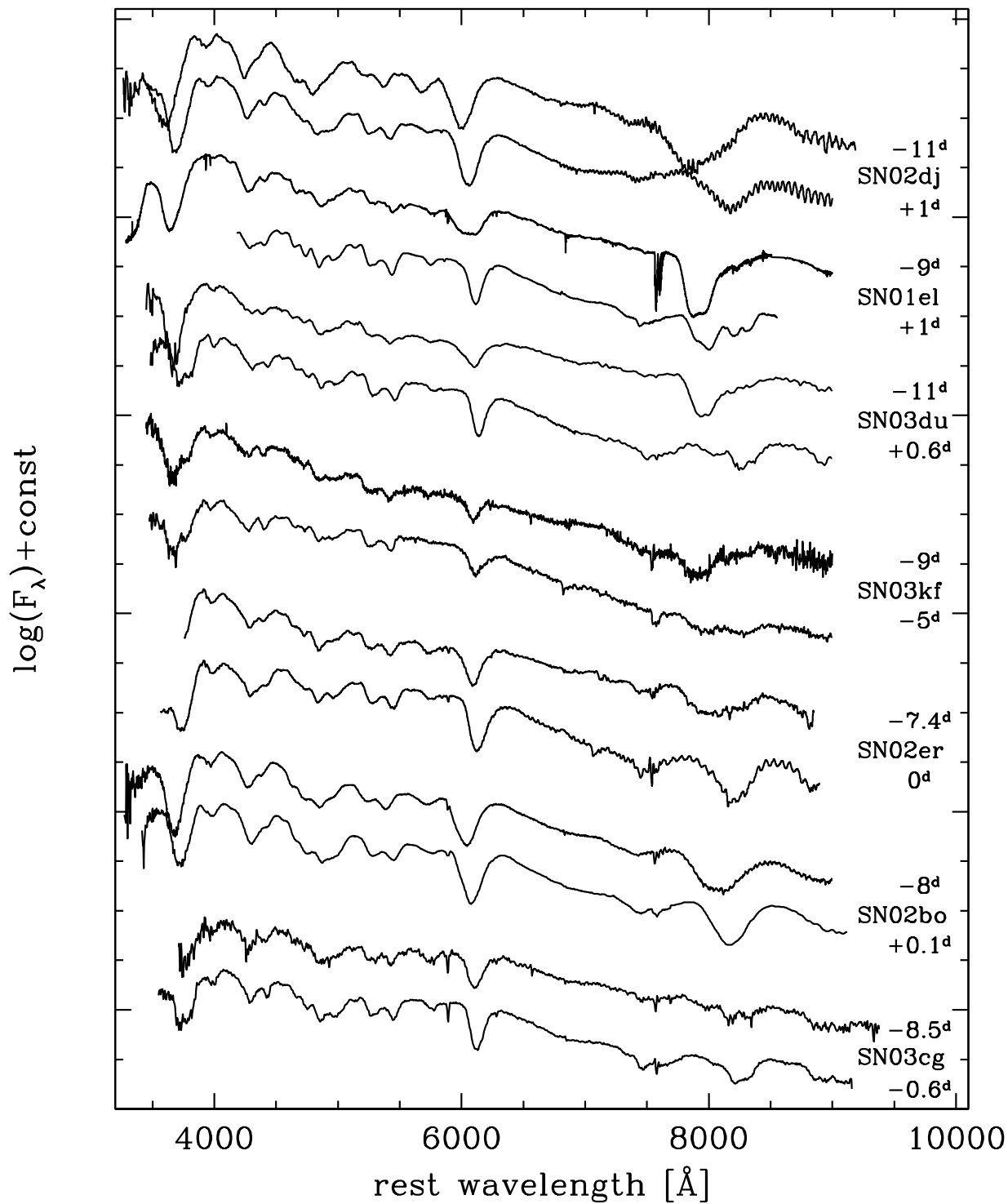}
\figcaption[early]{Spectra of the SNe~Ia observed by the RTN: for
each SN, the earliest spectrum is
shown, together with a spectrum near maximum. The spectra have been corrected
for reddening, arbitrarily scaled in flux, and are ordered by line velocity.
Atmospheric absorption was removed from all spectra except those of 
SN~2001el.}
\end{figure*}

\newpage
\begin{figure*}
\plotone{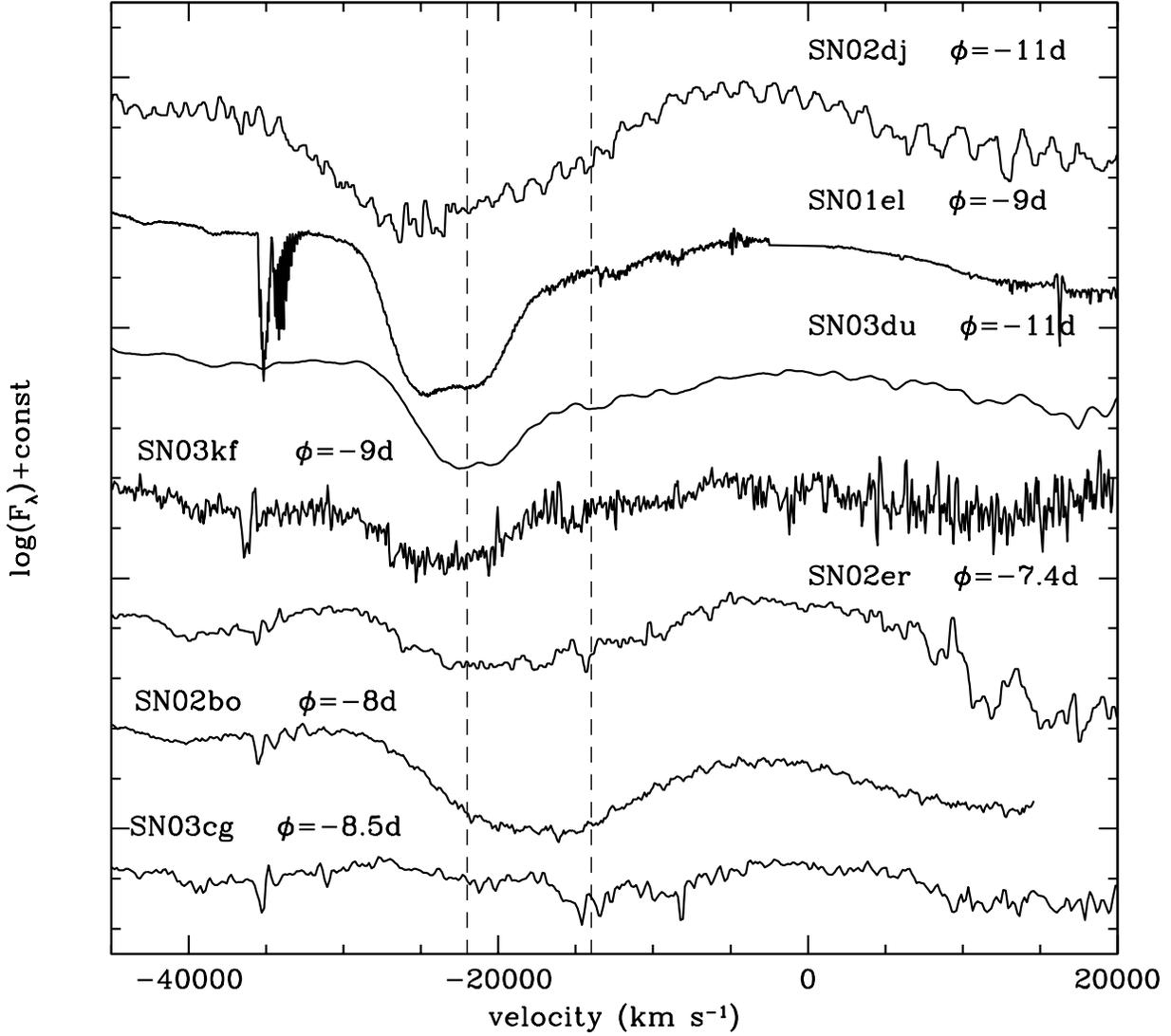}
\figcaption[CaII]{\CaII\ IR triplet profiles in the earliest spectra of the RTN
Supernovae. The abscissa shows velocity, taking the average wavelength of the
multiplet as the zero point, to emphasize the presence of the HVFs. The two
vertical lines mark the average position of the photospheric components 
(14000\,\kms) and of the HVFs (22000\,\kms), respectively.}
\end{figure*}


\end{document}